\documentclass[aps,preprint,tightenlines,superscriptaddress,showpacs]{revtex4}
\usepackage{times}
\usepackage{amsmath}
\usepackage{graphicx}
\usepackage{epsfig}
\usepackage{dcolumn}
\usepackage{bm}
\usepackage{color}
\newcommand{\BR}{{\cal B}}
\newcommand{\pp}{\pi^+\pi^-}
\newcommand{\pip}{\pi^+}
\newcommand{\pim}{\pi^-}

\newcommand{\ks}{K_S^0}
\newcommand{\piz}{\pi^0}

\newcommand{\kk}{K^+K^-}

\newcommand{\EE}{e^+e^-}

\newcommand{\etac}{\eta_c}
\newcommand{\jpsi}{J/\psi}
\newcommand{\pcpcjpsi}{\pi^+\pi^-J/\psi}

\newcommand{\ppjpsi}{\pcpcjpsi}
\newcommand{\hc}{h_c}
\newcommand{\pphc}{\pi^+\pi^-\hc}

\newcommand{\zc}{Z_c(3900)}
\newcommand{\zcp}{Z_c(4020)}

\def\Journal#1#2#3#4{{#1} {\bf #2}, #3 (#4)}

\def\PRL{Phys. Rev. Lett.}
\def\PRD{Phys. Rev. D}

\begin{document}

\title{
\boldmath
Observation of a charged charmoniumlike structure $\zcp$ and
search for the $\zc$ in $\EE\to \pp\hc$}

\author{
 {
M.~Ablikim$^{1}$, M.~N.~Achasov$^{8,a}$, O.~Albayrak$^{4}$,
D.~J.~Ambrose$^{41}$, F.~F.~An$^{1}$, Q.~An$^{42}$,
J.~Z.~Bai$^{1}$, R.~Baldini Ferroli$^{19A}$, Y.~Ban$^{28}$,
J.~Becker$^{3}$
% ...
\\
\vspace{0.2cm}
(BESIII Collaboration)\\
\vspace{0.2cm}
{\it
$^{1}$ Institute of High Energy Physics, Beijing 100049, People's Republic of China\\
% ...
$^{3}$ Bochum Ruhr-University, D-44780 Bochum, Germany\\
$^{4}$ Carnegie Mellon University, Pittsburgh, Pennsylvania 15213, USA\\
% ...
$^{8}$ G.I. Budker Institute of Nuclear Physics SB RAS (BINP), Novosibirsk 630090, Russia\\
% ...
$^{19}$ (A)INFN Laboratori Nazionali di Frascati, I-00044, Frascati, Italy; (B)INFN and University of Perugia, I-06100, Perugia, Italy\\
% ...
$^{28}$ Peking University, Beijing 100871, People's Republic of China\\
% ...
$^{41}$ University of Rochester, Rochester, New York 14627, USA\\
$^{42}$ University of Science and Technology of China, Hefei 230026, People's Republic of China\\
% ...
$^{a}$ Also at the Novosibirsk State University, Novosibirsk, 630090, Russia\\
% ...
}
\vspace{0.5cm}
}}

\vspace{0.5cm}
\date{\today}

\begin{abstract}

We study $\EE\to \pphc$ at center-of-mass energies from 3.90~GeV
to 4.42~GeV using data samples collected with the BESIII detector
operating at the Beijing Electron Positron Collider. The Born
cross sections are measured at 13 energies, and are found to be of
the same order of magnitude as those of $\EE\to \pcpcjpsi$ but
with a different line shape. In the $\pi^\pm \hc$ mass spectrum, a
distinct structure, referred to as $\zcp$, is observed at
4.02~GeV/$c^2$. The $\zcp$ carries an electric charge and couples
to charmonium. A fit to the $\pi^\pm\hc$ invariant mass spectrum,
neglecting possible interferences, results in a mass of
$(4022.9\pm 0.8\pm 2.7)~{\rm MeV}/c^2$ and a width of $(7.9\pm
2.7\pm 2.6)$~MeV for the $\zcp$, where the first errors are
statistical and the second systematic. The difference between the parameters
of this structure and the $Z_c(4025)$ observed in $D^{*}\bar{D}^{*}$ final
state is within $1.5 \sigma$, but whether they are the
same state needs further investigation. No significant $\zc$ signal
is observed, and upper limits on the $\zc$ production cross
sections in $\pi^\pm\hc$ at center-of-mass energies of 4.23 and
4.26~GeV are set.

\end{abstract}


\maketitle
\newpage

In the study of the $\EE\to \pcpcjpsi$ at center-of-mass (CM)
energies around 4.26~GeV, the BESIII~\cite{bes3_zc} and
Belle~\cite{belle_zc} experiments observed a charged
charmoniumlike state, the $\zc$, which was confirmed shortly after
with CLEO data at a CM energy of 4.17~GeV~\cite{seth_zc}. As there
are at least four quarks within the $\zc$, it is interpreted
either as a tetraquark state, $D\bar{D^*}$ molecule,
hadro-quarkonium, or other configuration~\cite{models}. More
recently, BESIII has observed another charged $Z_{c}(4025)$ state
in $\EE\to \pi^\pm(D^*\bar{D}^*)^\mp$~\cite{zc4025}. These states
together with similar states observed in the bottomonium
system~\cite{zb} would seem to indicate that a new class of
hadrons has been observed.

Such a particle may couple to $\pi^\pm\hc$~\cite{models} and thus
can be searched for in $\EE\to \pphc$. This final state has been
studied by CLEO~\cite{cleo_pphc},
and a hint of a rising cross section at 4.26~GeV has
been observed. An improved measurement may shed light on
understanding the nature of the $Y(4260)$ as
well~\cite{belle_y,babar_y}.

In this Letter, we present a study of $\EE\to \pphc$ at 13 CM
energies from 3.900 to 4.420~GeV. The data samples were collected
with the BESIII detector~\cite{bepc2}, and are listed in
Table~\ref{scan-data}. The CM energies ($\sqrt{s}$) are measured
with a beam energy measurement system~\cite{BEMS} with an
uncertainty of $\pm 1.0$~MeV. A charged structure is observed in
the $\pi^\pm\hc$ invariant mass spectrum at 4.02~GeV/$c^2$
(referred to as the $\zcp$ hereafter). We also report on the
search for $\zc$ decays into the same final state. No significant
signal is observed, and an upper limit on the production rate is
determined. In the studies presented here, the $\hc$ is
reconstructed via its electric-dipole (E1) transition $\hc\to
\gamma\etac$ with $\etac\to X_i$, where $X_i$ signifies 16
exclusive hadronic final states: $p \bar{p}$, $2(\pi^+ \pi^-)$,
$2(K^+ K^-)$, $K^+ K^- \pi^+ \pi^-$, $p \bar{p} \pi^+ \pi^-$,
$3(\pi^+ \pi^-)$, $K^+ K^- 2(\pi^+ \pi^-)$, $\ks K^\pm \pi^\mp$,
$\ks K^\pm \pi^\mp \pi^\pm \pi^\mp$, $K^+ K^- \pi^0$, $p
\bar{p}\pi^0$, $\pi^+ \pi^- \eta$, $K^+ K^- \eta$, $2(\pi^+ \pi^-)
\eta$, $\pi^+ \pi^- \pi^0 \pi^0$, and $2(\pi^+ \pi^-) \pi^0
\pi^0$. 

\begin{table}[htbp]
\caption{$\EE\to \pphc$ cross sections (or upper limits at the
90\% confidence level). The third errors are from the uncertainty
in $\BR(\hc\to \gamma\etac)$~\cite{bes3-hc-inclusive}.}
\label{scan-data}
\begin{tabular}{crcccc}
  \hline\hline
  $\sqrt{s}$~(GeV) & ${\cal L}$ (pb$^{-1}$)
  & $n^{\rm obs}_{\hc}$ & $\sigma(\EE\to \pphc)$~(pb) \\
  \hline
  3.900  &  52.8~~ & $<2.3$  & $<8.3 $ \\
  4.009  & 482.0~~ & $<13$  & $<5.0$ \\
  4.090  &  51.0~~ & $<6.0$  & $<13$ \\
  4.190  &  43.0~~ & $8.8\pm 4.9$  & $17.7\pm  9.8\pm  1.6\pm 2.8$ \\
  4.210  &  54.7~~ & $21.7\pm 5.9$ & $34.8\pm  9.5\pm  3.2\pm 5.5$ \\
  4.220  &  54.6~~ & $26.6\pm 6.8$ & $41.9\pm 10.7\pm  3.8\pm 6.6$ \\
  4.230  &1090.0~~ & $646\pm 33$   & $50.2\pm  2.7\pm  4.6\pm 7.9$ \\
  4.245  &  56.0~~ & $22.6\pm 7.1$ & $32.7\pm 10.3\pm  3.0\pm 5.1$ \\
  4.260  & 826.8~~ & $416\pm 28$   & $41.0\pm  2.8\pm  3.7\pm 6.4$ \\
  4.310  &  44.9~~ & $34.6\pm 7.2$ & $61.9\pm 12.9\pm  5.6\pm 9.7$ \\
  4.360  & 544.5~~ & $357\pm 25$   & $52.3\pm  3.7\pm  4.8\pm 8.2$ \\
  4.390  &  55.1~~ & $30.0\pm 7.8$ & $41.8\pm 10.8\pm  3.8\pm 6.6$ \\
  4.420  &  44.7~~ & $29.1\pm 7.3$ & $49.4\pm 12.4\pm  4.5\pm 7.6$ \\
  \hline\hline
\end{tabular}
\end{table}

We select charged tracks, photons, and $K^0_S\to \pi^+\pi^-$
candidates as described in Ref.~\cite{bes3-hc-exclusive}. A
candidate $\pi^0$~($\eta$) is reconstructed from pairs of photons
with an invariant mass in the range $|M_{\gamma\gamma}-m_{\pi^0}|
< 15~\hbox{\rm MeV}/c^2$ ($|M_{\gamma\gamma}-m_{\eta}| <
15~\hbox{\rm MeV}/c^2$), where $m_{\pi^0}$ ($m_{\eta}$) is the
nominal $\pi^0$~($\eta$) mass~\cite{PDG}.

In selecting $\EE\to \pphc$, $\hc\to \gamma\etac$ candidates, all
charged tracks are assumed to be pions, and events with at least
one combination satisfying $M_{\pp}^{\rm recoil}\in
[3.45,~3.65]$~GeV/$c^2$ and $M_{\gamma\pp}^{\rm recoil}\in
[2.8,~3.2]$~GeV/$c^2$ are kept for a further analysis. Here
$M_{\pp}^{\rm recoil}$ ($M_{\gamma\pp}^{\rm recoil}$) 
is the mass recoiling from the $\pp$ ($\gamma\pp$) pair,
which should be in the mass range of  the $\hc$ ($\etac$). 

To determine the species of final state particles and to select
the best photon when additional photons (and $\piz$ or $\eta$
candidates) are found in an event, the combination with the
minimum value of $\chi^{2}=\chi^{2}_{\rm 4C} +
\sum_{i=1}^{N}\chi^{2}_{\rm PID}(i)+\chi^{2}_{\rm 1C}$ is selected
for a further analysis, where $\chi^{2}_{\rm 4C}$ is the $\chi^2$
from the initial-final four-momentum conservation (4C) kinematic
fit, $\chi^{2}_{\rm PID}(i)
$ is the $\chi^{2}$ from particle identification using the energy loss
in the MDC and the time measured with the Time-of-Flight system.
$N$ is the number of the charged tracks in the final states, and
$\chi^{2}_{\rm 1C}$ is the sum of the 1C (mass constraint of the
two daughter photons) $\chi^{2}$ of the $\piz$ and $\eta$ in each
final state. There is also a $\chi^{2}_{\rm 4C}$ requirement,
which is optimized using the figure-of-merit, $S/\sqrt{S+B}$,
where $S$ and $B$ are the numbers of MC simulated signal and
background events, respectively, and $\chi^{2}_{\rm 4C}<35$
(efficiency is about 80\% from MC simulation) is required for
final states with only charged or $\ks$ particles, while
$\chi^{2}_{\rm 4C}<20$ (efficiency is about 70\% from MC
simulation) is required for those with $\piz$ or $\eta$~\cite{4C}.
A similar optimization procedure determines the $\etac$ candidate
mass window around the nominal $\etac$~\cite{PDG} mass to be $\pm
50$~MeV/$c^2$ with efficiency about 85\% from MC simulation ($\pm
45$~MeV/$c^2$ with efficiency about 80\% from MC simulation) for
final states with only charged or $\ks$ particles (those with
$\piz$ or $\eta$).

\begin{figure}[htbp]
\begin{center}
\includegraphics[width=0.68\textwidth]{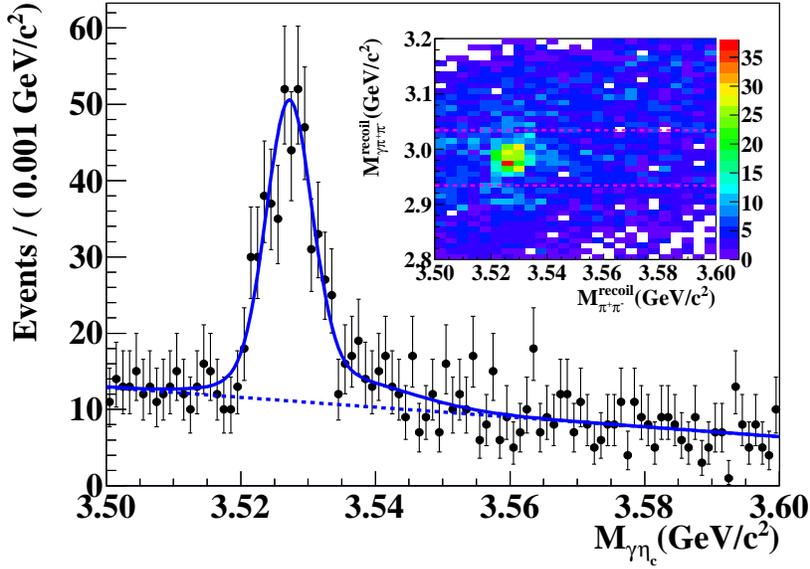}
\caption{
The $M_{\gamma\etac}$ distribution after the $\etac$ signal
selection of 4.26~GeV data, dots with error bars are data and the
curves are the best fit described in the text. 
The inset is the scatter plot of the mass of the 
$\etac$ candidate versus that of the $\hc$ candidate.} \label{scatter}
\end{center}
\end{figure}

Figure~\ref{scatter} shows as an example the scatter plot of the
mass of the $\etac$ candidate versus that of the $\hc$ candidate
at the CM energy of 4.26~GeV, as well as the projection of the
invariant mass distribution of $\gamma\etac$ in the $\etac$ signal
region, where a clear $\hc\to \gamma\etac$ signal is observed. To
extract the number of $\pphc$ signal events, the $\gamma\etac$
mass spectrum is fitted using the MC simulated signal shape
convolved with a Gaussian function to reflect the mass resolution
difference (around 10\%) between data and MC simulation, together
with a linear background. The fit to the 4.26~GeV data is shown in
Fig.~\ref{scatter}. The tail in the high
mass side is due to the events with initial state radiation (ISR)
which is simulated well in MC, and its fraction is fixed in the
fit. At the energy points with large statistics (4.23, 4.26, and
4.36~GeV), the fit is applied to the 16 $\etac$ decay modes
simultaneously, while at the other energy points, we fit the mass
spectrum summed over all the $\etac$ decay modes. The number of
signal events ($n^{\rm obs}_{\hc}$) and the measured Born cross
section at each energy are listed in Table~\ref{scan-data}. The
$\pphc$ cross section appears to be constant above 4.2~GeV with a
possible local maximum at around 4.23~GeV. This is in contrast to
the observed energy dependence in the $\EE\to \ppjpsi$ channel
which revealed a decrease of cross sections at higher
energies~\cite{belle_zc,babarnew}.

Systematic errors in the cross section measurement mainly come
from the luminosity measurement, the branching fraction of $\hc\to
\gamma\etac$, the branching fraction of $\etac\to X_i$, the
detection efficiency, the ISR correction factor, and the fit. The
integrated luminosity at each energy point is measured using large
angle Bhabha events, and it has an estimated uncertainty of 1.2\%.
The branching fractions of $\hc\to \gamma\etac$ and $\etac\to
X_{i}$ are taken from
Refs.~\cite{bes3-hc-exclusive,bes3-hc-inclusive}. The
uncertainties in the detection efficiency are estimated in the
same way as described in Refs.~\cite{bes3-hc-exclusive,guoyp}, and
the error in the ISR correction is estimated as described in
Ref.~\cite{bes3_zc}. Uncertainties due to the choice of the signal
shape, the background shape, the mass resolution, and fit range
are estimated by varying the $\hc$ and $\etac$ resonant parameters
and line shapes in MC simulation, varying the background function
from linear to a second-order polynomial, varying the mass
resolution difference between data and MC simulation by one
standard deviation, and by extending the fit range. Assuming all
of the sources are independent, the total systematic error in the
$\pphc$ cross section measurement is determined to be between 7\%
and 9\% depending on the energy, and to be conservative we take
9\% for all the energy points. The uncertainty in
$\BR(\hc\to\gamma\etac)$ is 15.7\%~\cite{PDG}, common to all
energy points, and quoted separately in the cross section
measurement. Altogether, about 95\% of the total systematic errors
are common to all the energy points.

\begin{figure}[htbp]
\begin{center}
\includegraphics[width=0.68\textwidth]{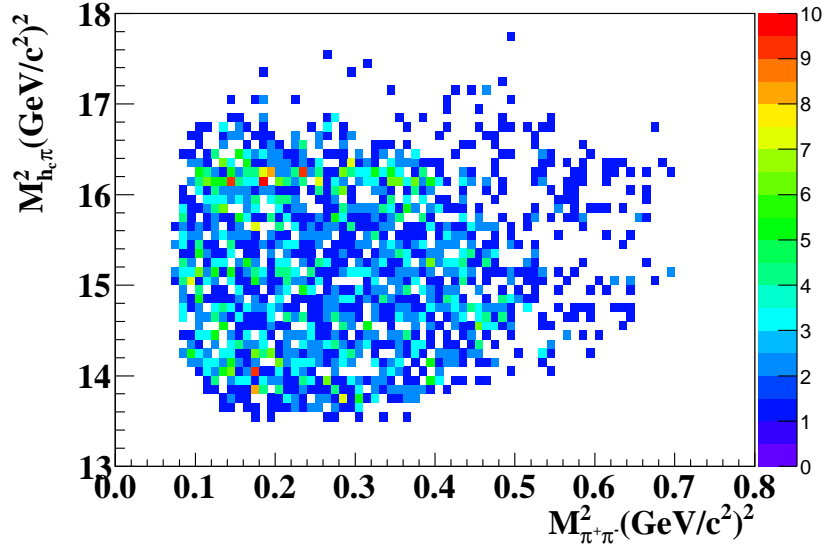}
\caption{Dalitz plot ($M^2_{\pi^+\hc}$ vs. $M^2_{\pp}$) for
selected $\EE\to \pphc$ events, summed over all energy points.}
\label{dalitz}
\end{center}
\end{figure}

\begin{figure}[htbp]
\begin{center}
\includegraphics[width=0.68\textwidth]{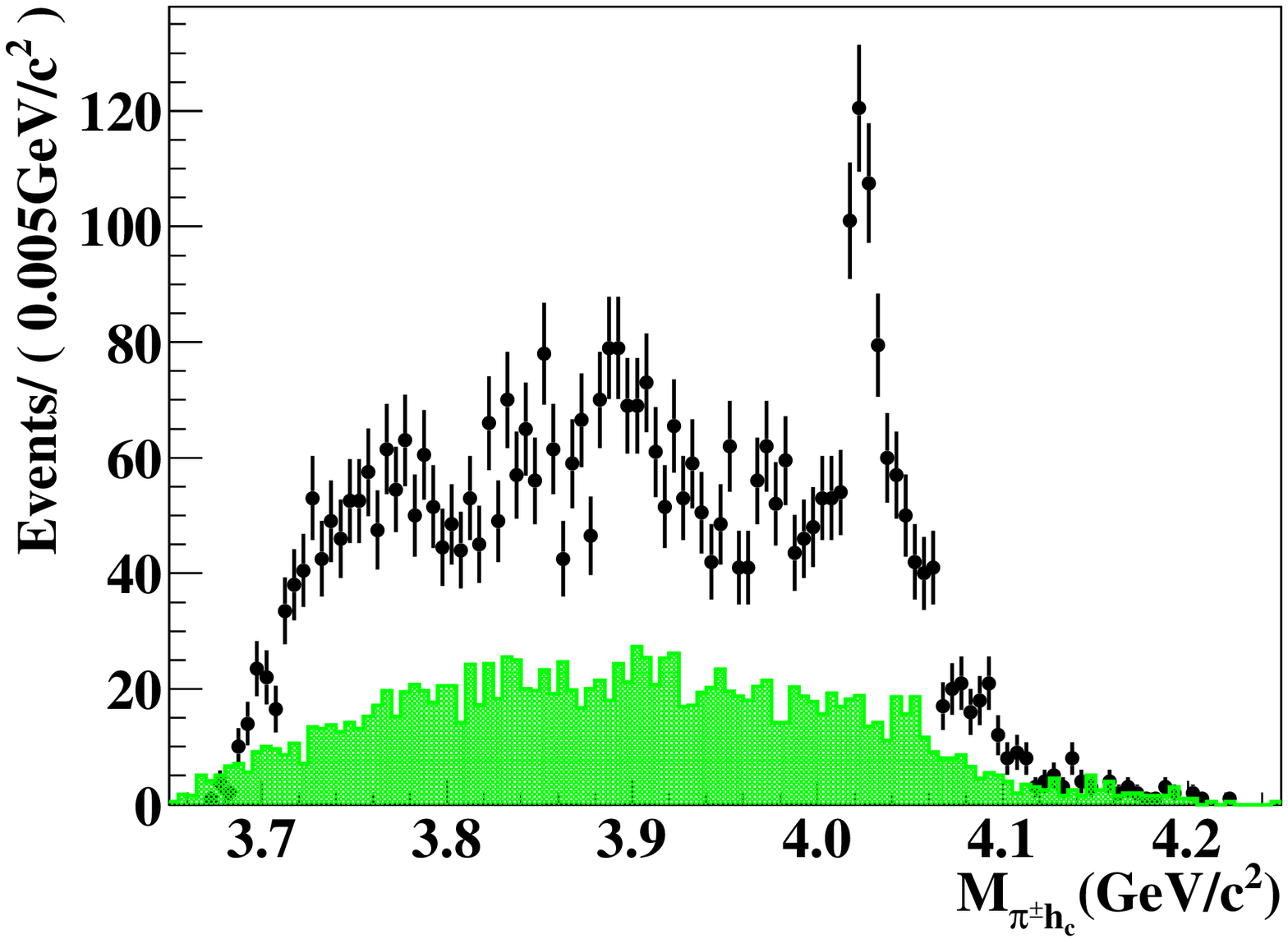}
\caption{$M_{\pi^\pm\hc}$ distribution of $\EE\to \pphc$ candidate
events in the $\hc$ signal region (dots with error bars) and the
normalized $\hc$ sideband region (shaded histogram), summed over
data at all energy points.} \label{proj}
\end{center}
\end{figure}

Intermediate states are studied by examining the Dalitz plot of
the selected $\pphc$ candidate events. The $\hc$ signal is
selected using $3.518 < M_{\gamma \etac} < 3.538$~GeV/$c^2$ and
the sideband using $3.490 < M_{\gamma \etac} < 3.510$~GeV/$c^2$ or
$3.560 < M_{\gamma \etac} < 3.580$~GeV/$c^2$, which is twice as
wide as the signal region. 
Figure~\ref{dalitz} shows the
Dalitz plot of the $\pphc$ candidate events summed over all
energies. While there are no clear structures in the $\pp$ system,
there is clear evidence for an exotic charmoniumlike structure in
the $\pi^\pm\hc$ system. Figure~\ref{proj} shows the projection of
the $M_{\pi^\pm\hc}$ (two entries per event) distribution for the
signal events, as well as the background events estimated from
normalized $\hc$ mass sidebands. There is a significant peak at
around 4.02~GeV/$c^2$ (the $\zcp$), and the wider peak at low
masses is the reflection of the $\zcp$. There are also some events
at around 3.9~GeV/$c^2$, which could be the $\zc$.  The individual
data sets at 4.23~GeV, 4.26~GeV and 4.36~GeV show similar
structures.

\begin{figure}[htbp]
\begin{center}
\includegraphics[width=0.68\textwidth]{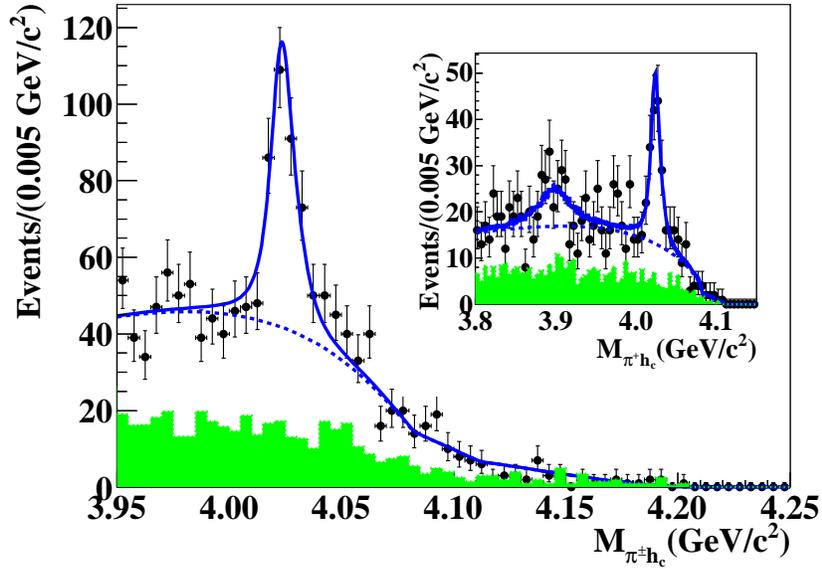}
\caption{Sum of the simultaneous fits to the $M_{\pi^\pm\hc}$
distributions at 4.23~GeV, 4.26~GeV, and 4.36~GeV as described in
the text; the inset shows the sum of the simultaneous fit to the
$M_{\pi^+\hc}$ distributions at 4.23~GeV and 4.26~GeV with $\zc$
and $\zcp$. Dots with error bars are data; shaded histograms are
normalized sideband background; the solid curves show the total
fit, and the dotted curves the backgrounds from the fit.}
\label{1Dfit}
\end{center}
\end{figure}

An unbinned maximum likelihood fit is applied to the
$M_{\pi^\pm\hc}$ distribution summed over the 16 $\etac$ decay
modes. The data at 4.23~GeV, 4.26~GeV, and 4.36~GeV are fitted
simultaneously with the same signal function with common mass and
width. The signal shape is parameterized as a constant width
relativistic Breit-Wigner~(BW) function convolved with a Gaussian
with a mass resolution determined from data directly. Assuming the
spin-parity of the $\zcp$ $J^P=1^+$, a phase space factor $pq^3$
is considered in the partial width, where $p$ is the $\zcp$
momentum in the $\EE$ CM frame and $q$ is the $\hc$ momentum in
the $\zcp$ CM frame. The background shape is parameterized as an
ARGUS function~\cite{ARGUS}. The efficiency curve is considered in
the fit, but possible interferences between the signal and
background are neglected. Figure~\ref{1Dfit} shows the fit
results; the fit yields a mass of $(4022.9\pm 0.8)~{\rm MeV}/c^2$,
and a width of $(7.9\pm 2.7)$~MeV. The goodness-of-fit is found to
be $\chi^{2}/ndf=27.3/32=0.85$
 by projecting the events into a histogram with 46 bins.
The statistical significance of the $\zcp$ signal is calculated by
comparing the fit likelihoods with and without the signal. Besides
the nominal fit, the fit is also performed by changing the fit
range, the signal shape, or the background shape. In all cases,
the significance is found to be greater than $8.9\sigma$.

The numbers of $\zcp$ events are determined to be $N(\zcp^\pm) =
114\pm 25$, $72\pm 17$, and $67\pm 15$ at 4.23~GeV, 4.26~GeV, and
4.36~GeV, respectively. The cross sections are calculated to be
$\sigma(\EE\to \pi^\pm \zcp^\mp\to \pphc) = (8.7\pm 1.9\pm 2.8\pm
1.4)$~pb at 4.23~GeV, $(7.4\pm 1.7\pm 2.1\pm 1.2)$~pb at 4.26~GeV,
and $(10.3\pm 2.3\pm 3.1\pm 1.6)$~pb at 4.36~GeV, where the first
errors are statistical, the second ones systematic (described in
detail below), and the third ones from the uncertainty in
$\BR(\hc\to \gamma\etac)$~\cite{PDG}. The $\zcp$ production rate
is uniform at these three energy points.

Adding a $\zc$ with mass and width fixed to the BESIII
measurement~\cite{bes3_zc} in the fit, results in a statistical
significance of 2.1$\sigma$ (see the inset of Fig.~\ref{1Dfit}).
We set upper limits on the production cross sections as
$\sigma(\EE\to \pi^\pm \zc^\mp\to \pphc) <13$~pb at 4.23~GeV and
$<11$~pb at 4.26~GeV, at the 90\% confidence level (C.L.). The
probability density function from the fit is smeared by a Gaussian
function with standard deviation of $\sigma_{\rm sys}$ to include
the systematic error effect, where $\sigma_{\rm sys}$ is the
relative systematic error in the cross section measurement
described below. We do not fit the 4.36~GeV data as the $\zc$
signal overlaps with the reflection of the $\zcp$ signal.

The systematic errors for the resonance parameters of the $\zcp$
come from the mass calibration, parametrization of the signal and
background shapes, possible existence of the $\zc$ and
interference with it, fitting range, efficiency curve, and the
mass resolution. The uncertainty from the mass calibration is
estimated using the difference between the measured and known
$\hc$ masses and $D^0$ masses (reconstructed from $K^-\pip$). The
differences are $(2.1\pm 0.4)$~MeV/$c^2$ and $-(0.7\pm
0.2)$~MeV/$c^2$, respectively. Since our signal topology has one
low momentum pion and many tracks from the $\hc$ decay, we assume
these differences added in quadrature, 2.6~MeV/$c^2$, is the
systematic error 
due to the mass
calibration. Spin-parity conservation forbids a zero spin for the
$\zcp$, and assuming that contributions from D-wave or higher are
negligible, the only alternative is $J^P=1^-$ for the $\zcp$. 
A fit under this scenario yields a mass difference of
0.2~MeV/$c^2$ and a width difference of 0.8~MeV. The uncertainty
due to the background shape is determined by changing to a
second-order polynomial and by varying the fit range. A difference
of 0.1~MeV/$c^2$ for the mass is found from the former, and
differences of 0.2~MeV/$c^2$ for mass and 1.1~MeV for width are
found from the latter. Uncertainties due to the mass resolution
are estimated by varying the resolution difference between data
and MC simulation by one standard deviation of the measured
uncertainty in the mass resolution of the $\hc$ signal; the
difference is 0.5~MeV in the width, which is taken as the
systematic error. The uncertainty in the efficiency curve results
in 0.1~MeV/$c^2$ for mass and 0.1~MeV for width. Uncertainties due
to the possible existence of the $\zc$ and the interference with
it are estimated by adding a $\zc$ amplitude incoherently or
coherently in the fit. The uncertainties due to $\zc$ is
0.2~MeV/$c^{2}$ for mass and 2.1~MeV for width, while the
uncertainties due to interference is 0.5~MeV/$c^{2}$ for the mass
and 0.4~MeV for the width. Assuming all the sources of systematic
uncertainty are independent, the total systematic error is
2.7~MeV/$c^2$ for the mass, and 2.6~MeV for the width.

The systematic errors in $\sigma(\EE\to \pi^\pm \zcp^\mp\to
\pphc)$ are estimated in the same way as for $\sigma(\EE\to
\pphc)$. The systematic errors due to the inclusion of the $\zc$
signal, the possible interference between $\zcp$ and $\zc$, the
fitting range, the signal and background parameterizations,
the $\hc$ signal window selection, the mass resolution, and
the efficiency curve, in addition to those in the
$\sigma(\EE\to \pphc)$ measurement, are considered and summarized
in Table~\ref{sys}. The systematic errors in $\sigma(\EE\to
\pi^\pm \zc^\mp\to \pphc)$ are determined similarly.

\begin{table*}[htbp]
\caption{The percentage systematic errors in $\sigma(\EE\to
\pi^\pm \zcp^\mp\to \pphc)$, in addition to those in
$\sigma(\EE\to \pphc)$ measurement.} \label{sys}
 {\scriptsize
\begin{tabular}{c|c|c|c|c|c|c|c|c}
  \hline\hline
  $\sqrt{s}$~(GeV) & $\zc$ signal  & interference & fitting range
        & signal shape & background shape & $\hc$ signal window
        & mass resolution  &  efficiency curve\\
\hline
 4.230   &18.3   & 20.0  & 13.2    & 4.5     & 3.5    & 1.7 & 1.8  & 0.9\\
 4.260   &16.2   & 20.0  & 8.3     & 4.2     & 2.8    & 1.7  & 1.8   & 0.0\\
 4.360   &18.3   & 20.0  & 4.5     & 6.0     & 6.0    & 1.4  & 1.5   & 0.0 \\
 \hline\hline
\end{tabular}
 }
\end{table*}

In summary, we measure $\EE\to \pphc$ cross sections at CM
energies between 3.90 and 4.42~GeV for the first time. These cross
sections are of the same order of magnitude as those of the
$\EE\to \pcpcjpsi$ measured by BESIII~\cite{bes3_zc} and other
experiments~\cite{belle_zc,babarnew}, but with a different line
shape. There is a broad structure at high energy with a possible
local maximum at around 4.23~GeV. A narrow structure very close to
the $(D^\ast\bar{D}^\ast)^\pm$ threshold with a mass of
$(4022.9\pm 0.8\pm 2.7)~{\rm MeV}/c^2$ and a width of $(7.9\pm
2.7\pm 2.6)$~MeV is observed in the $\pi^\pm \hc$ mass spectrum.
This structure couples to charmonium and has an electric charge,
which is suggestive of a state containing more quarks than just a
charm and an anti-charm quark, as the $\zc$ observed in the
$\pi^\pm\jpsi$ system~\cite{bes3_zc,belle_zc,seth_zc}. We do not
find a significant signal for  $\zc\to\pi^\pm\hc$ and the
production cross section is found to be smaller than 11~pb at the
90\% C.L. at 4.26~GeV, which is lower than that of $\zc\to
\pi^\pm\jpsi$~\cite{bes3_zc}. The $\zcp$ parameters agree within
1.5$\sigma$ of those of the $Z_c(4025)$, observed in $\EE\to
\pi^\pm (D^*\bar{D}^*)^\mp$ at CM energy 4.26~GeV ~\cite{zc4025}.
Results for the latter at 4.23 and 4.36~GeV may help us to
understand whether they are the same state.


The BESIII collaboration thanks the staff of BEPCII and the
computing center for their strong support. This work is supported
in part by the Ministry of Science and Technology of China under
Contract No. 2009CB825200; National Natural Science Foundation of
China (NSFC) under Contracts Nos. 10625524, 10821063, 10825524,
10835001, 10935007, 11125525, 11235011; Joint Funds of the
National Natural Science Foundation of China under Contracts Nos.
11079008, 11079023, 11179007, U1332201; the Chinese Academy of Sciences (CAS)
Large-Scale Scientific Facility Program; CAS under Contracts Nos.
KJCX2-YW-N29, KJCX2-YW-N45; 100 Talents Program of CAS; German
Research Foundation DFG under Contract No. Collaborative Research
Center CRC-1044; Istituto Nazionale di Fisica Nucleare, Italy;
Ministry of Development of Turkey under Contract No.
DPT2006K-120470; U. S. Department of Energy under Contracts Nos.
DE-FG02-04ER41291, DE-FG02-05ER41374, DE-FG02-94ER40823; U.S.
National Science Foundation; University of Groningen (RuG) and the
Helmholtzzentrum fuer Schwerionenforschung GmbH (GSI), Darmstadt;
WCU Program of National Research Foundation of Korea under
Contract No. R32-2008-000-10155-0.



\begin{thebibliography}{**}
\bibitem{4C}
A track parameter correction procedure~\cite{guoyp} is
applied to the MC simulated charged tracks to improve the
agreement between data and MC simulation of the $\chi^2_{\rm 4C}$
distribution of the kinematic fit. The correction factors are
obtained from a high purity control sample $\jpsi\to \phi
f_{0}(980)$, with $\phi\to \kk$ and $f_{0}(980)\to \pip\pim$.
Reasonable agreement between data and MC simulation is observed
for the channels analyzed in this work. The systematic error of
the $\chi^{2}_{\rm 4C}$ requirement is taken as half of the
correction in the efficiency, which is much larger than the effect
due to the uncertainties in the correction factors to cover
possible unknown effects in the procedure.
\end{thebibliography}
\end{document}